\documentclass[trackchanges, twocolumn, longbib]{aastex701}

\usepackage{amsmath,amssymb}
\usepackage{comment}

\newcommand{\mgnecola}{\texttt{MG-NECOLA}}
\newcommand{\mgcola}{\texttt{MG-PICOLA}}
\newcommand{\mgquijote}{\texttt{QUIJOTE-MG}}
\newcommand{\lcdm}{$\Lambda$CDM}
\newcommand{\N}{\text{N}}
\newcommand{\eff}{\text{eff}}
\newcommand{\Mpc}{\text{Mpc}}
\newcommand{\m}{\text{m}}

\definecolor{darkgreen}{rgb}{0, 0.6, 0}

\begin{document}

\title{MG-NECOLA: A Field-Level Emulator for $f(R)$ Gravity and Massive Neutrino Cosmologies}

\author[orcid=0000-0001-5090-2860]{J. Bayron Orjuela-Quintana}
\affiliation{Departamento  de  F\'isica,  Universidad  del Valle, Ciudad  Universitaria Mel\'endez, Santiago de Cali  760032,  Colombia}
\affiliation{Department of Physics, Michigan Technological University, Houghton, MI 49931, USA}
\email[show]{john.orjuela@correounivalle.edu.co}
\correspondingauthor{J. Bayron Orjuela-Quintana}

\author[orcid=0009-0005-5084-074X]{Mauricio Reyes}
\affiliation{Department of Physics, Michigan Technological University, Houghton, MI 49931, USA}
\email{mrhurtad@mtu.edu}

\author[orcid=0000-0003-3052-3059]{Elena Giusarma}
\affiliation{Department of Physics, Michigan Technological University, Houghton, MI 49931, USA}
\email{egiusarm@mtu.edu}

\author[orcid=0000-0003-4145-1943]{Marco Baldi}
\affiliation{Dipartimento di Fisica e Astronomia ``Augusto Righi'', Università di Bologna, via Piero Gobetti 93/2, I-40129 Bologna, Italy}
\affiliation{INAF, Osservatorio di Astrofisica e Scienza dello Spazio di Bologna, via Piero Gobetti 93/3, I-40129 Bologna, Italy}
\email{marco.baldi5@unibo.it}

\author[orcid=0000-0003-4786-2348]{Neerav Kaushal}
\affiliation{Department of Applied Computing, Michigan Technological University, Houghton, MI 49931, USA}

\email{kaushal@mtu.edu}

\author[orcid=0000-0002-6380-2778]{C\'esar A. Valenzuela-Toledo}
\affiliation{Departamento  de  F\'isica,  Universidad  del Valle, Ciudad  Universitaria Mel\'endez,  Santiago de Cali  760032,  Colombia}
\email{cesar.valenzuela@correounivalle.edu.co}

\begin{abstract}
Accurate modeling of non-linear gravitational dynamics is essential for constraining extensions to the standard cosmological model using large-scale structure observations. While high-resolution $N$-body simulations provide the required fidelity, they are computationally prohibitive for the large ensembles needed to analyze Modified Gravity (MG) scenarios. We present \mgnecola, a field-level emulator based on a convolutional neural network that upgrades fast, approximate \mgcola~simulations to near--$N$-body accuracy at a fraction of the computational cost. Trained on a suite of \mgquijote~simulations for $f(R)$ gravity, \mgnecola~achieves nearly sub-percent accuracy ($\lesssim 1\%$) in both the matter power spectrum and bispectrum up to $k \simeq 1~h\,\mathrm{Mpc}^{-1}$. Crucially, although being trained on a fixed cosmology, the network generalizes robustly to cosmologies outside its training manifold keeping the error below 5\%. It successfully recovers the General Relativity limit (\lcdm) without introducing spurious MG signals and accurately captures the power suppression induced by massive neutrinos ($M_\nu \leq 0.4$ eV), despite being trained on cosmologies with massless neutrinos. The pipeline delivers a speed-up factor of $\sim 1500\times$ relative to full $N$-body runs, generating a high-fidelity realization in $\mathcal{O}(10^3)$ CPU seconds compared to $\mathcal{O}(10^6)$ for the baseline. This accuracy–efficiency trade-off establishes \mgnecola~as a powerful tool for generating the massive mock catalogs required for next-generation galaxy surveys.
\end{abstract}

\section{Introduction}

The prevailing theoretical framework of modern cosmology, known as the $\Lambda$ Cold Dark Matter (\lcdm) model, combines General Relativity (GR) with a Universe dominated by cold dark matter (CDM) and dark energy. While \lcdm~has been remarkably successful in explaining a wide range of observations, growing evidence points to possible inconsistencies between different cosmological probes~\citep{Riess:2021jrx, Planck:2018vyg, Hildebrandt:2018yau, Joudaki:2019pmv, Abdalla:2022yfr, DiValentino:2020vvd, DESI:2025zgx}. These observational tensions, together with the fact that GR has never been directly tested on the largest cosmic scales, provide strong motivation to explore extensions of the standard model in the form of \textit{modified gravity} (MG) theories~\citep{Clifton:2011jh, CANTATA:2021asi}. Such models introduce additional degrees of freedom that modify the growth of cosmic structures, potentially enhancing or suppressing gravitational interactions on cosmological scales.

Current and upcoming galaxy surveys will deliver unprecedentedly precise measurements of the matter distribution in the Universe, offering a unique opportunity to probe gravity beyond GR~\citep{DESI:2025zgx, Euclid:2024yrr, LSST:2019, WFIRST:2015}. Statistical observables such as the matter power spectrum (MPS), correlation functions, and higher-order statistics encode the imprint of gravitational physics on the Large-Scale Structure (LSS). Fully exploiting the constraining power of these data, however, requires fast and accurate theoretical predictions across a wide range of cosmological and gravitational parameters.

High-resolution $N$-body simulations provide the most reliable predictions for non-linear structure formation~\citep{Springel:2005mi, Springel:2020plp, Teyssier:2001cp, Potter:2016ttn}, but their computational cost makes them impractical for exploring the high-dimensional parameter spaces characteristic of MG cosmologies. Approximate methods, such as the \textit{COmoving Lagrangian Acceleration} (COLA) approach~\citep{Tassev:2013pn, Howlett:2015hfa}, offer a much more efficient alternative and accurately reproduce large-scale clustering. However, their accuracy deteriorates on mildly and fully non-linear scales, precisely where MG effects and screening mechanisms become most relevant.

To bridge this gap, recent efforts have turned to machine learning to accelerate LSS predictions. Among several strategies, one avenue involves the derivation of explicit analytical or semi-analytical approximations for the power spectrum. Recent works have employed symbolic regression and genetic algorithms to discover compact mathematical formulas for the linear power spectrum in both \lcdm~and extended scenarios~\citep{Orjuela-Quintana:2022nnq, Orjuela-Quintana:2023uqb, Orjuela-Quintana:2024hha, Bartlett:2023cyr, Bartlett:2024jes, Sui:2024wob}. These expressions provide an incredibly fast alternative to Boltzmann solvers~\citep{Blas:2011, Lewis:1999bs} and can be coupled with non-linear mappings like \texttt{Halofit}~\citep{Smith:2003, Takahashi:2012em} to estimate clustering. However, while efficient for summary statistics, these approaches do not provide full three-dimensional realizations of the matter distribution.

To address the need for field-level predictions, recent efforts have turned to other approaches. For instance, \cite{Berger:2018aey, He:2018ggn, AlvesdeOliveira:2020yix, Giusarma:2019feb, Jamieson:2022lqc, Bartlett:2024cra, Kaushal:2021hqv} have demonstrated the potential of neural networks to model non-linear structure formation. However, extending these generative methods to MG scenarios remains a challenge, as the network must learn not only the non-linear clustering of matter but also the environment-dependent screening mechanisms inherent to scalar-tensor theories. A recent attempt to mitigate this issue has been proposed by~\cite{Saadeh:2024vuj} where \lcdm~$N$-body simulations are turned into one evolved considering MG effects.

In this work, we go a step further to address this challenge by introducing \mgnecola, a deep learning framework based on convolutional neural networks (CNNs) that operates at the field level, correcting fast approximate \mgcola~simulations~\citep{Wright:2017dkw, Winther:2017jof} to near–$N$-body fidelity. Rather than emulating observables from scratch, \mgnecola~learns the non-linear displacement residuals between \mgcola~outputs and full $N$-body simulations from the \mgquijote~suite~\citep{mgquijote, Villaescusa-Navarro:2019bje}. This strategy preserves the large-scale accuracy of the underlying approximate solver while efficiently correcting the particle positions on small scales.

We demonstrate that \mgnecola~accurately reproduces key summary statistics, including the MPS and bispectrum, up to $k \simeq 1~h\,\mathrm{Mpc}^{-1}$, while remaining orders of magnitude faster than full $N$-body simulations. Crucially, despite being trained on a specific suite of MG simulations with massless neutrinos, the method generalizes robustly to unseen physical regimes. It successfully recovers the \lcdm~limit and accurately captures the power suppression in massive neutrino cosmologies. By producing full three-dimensional realizations at a fraction of the computational cost, \mgnecola~enables the efficient generation of large ensembles of high-fidelity simulations, providing a scalable tool for testing beyond-\lcdm~physics.\footnote{Preliminary results and the core architecture of this framework were presented in the NeurIPS Workshop on Machine Learning and the Physical Sciences~\citep{Orjuela-Quintana:2025wne}.}

This paper is organized as follows. In Sec.~\ref{Sec: Effects of MG} we review the theoretical framework of $f(R)$ gravity and its impact on LSS formation. In Sec.~\ref{Sec: Method} we describe the simulation data and the architecture of the neural network. Section~\ref{Sec: Results} presents the main performance results and validation against $N$-body simulations. Finally, in Section~\ref{Sec: Conclusions} we discuss the implications, limitations, and future extensions of our approach.

\section{Effect of Modified Gravity on Matter Clustering}
\label{Sec: Effects of MG}

In the standard \lcdm~framework, the growth of cosmic structures is governed by GR, where metric perturbations are sourced by matter density fluctuations through the Einstein equations. MG theories extend this picture by introducing additional degrees of freedom that alter the gravitational interaction on cosmological scales, leading to potentially enhanced or suppressed structure growth. These deviations are particularly relevant in the non-linear regime of LSS formation, where screening mechanisms and environment-dependent effects become dominant~\citep{Clifton:2011jh, CANTATA:2021asi}.

On sub-horizon scales and under the quasi-static approximation, the impact of MG on matter clustering can be effectively described by a scale- and time-dependent modification of the Poisson equation:
\begin{equation}
\label{Eq: Poisson Equation}
    \nabla^2 \Phi = 4\pi G_\eff(k, a)\, a^2 \bar{\rho}_\m \delta_\m,
\end{equation}
where $\Phi$ is the gravitational potential, $\bar{\rho}_\m$ is the mean matter density, $\delta_\m$ is the matter overdensity, and the effective gravitational coupling $G_\eff(k,a)$ encapsulates departures from GR. Unlike standard gravity, where linear growth is scale-independent, MG models generically induce a scale-dependent growth rate that propagates into the MPS and higher-order statistics.

Among MG models, $f(R)$ gravity provides a well-motivated and widely used benchmark. In these theories, the Einstein–Hilbert action is generalized by a non-linear function of the Ricci scalar, introducing an additional scalar degree of freedom—the scalaron—that mediates a fifth force~\citep{Sotiriou:2008rp, DeFelice:2010aj}. In the linear regime, this leads to an effective gravitational coupling of the form~\citep{Cardona:2022lcz, Orjuela-Quintana:2023zjm}:
\begin{equation}
    \frac{G_\text{eff}(k, a)}{G_\text{N}} = 1 + \frac{1}{3} \frac{k^2}{k^2 + a^2 m^2(a)},
\end{equation}
where $G_\text{N}$ is Newton's constant and $m(a)$ is the mass of the scalaron, which governs the range of the fifth force. Gravity is enhanced by a factor of $4/3$ on scales smaller than the scalaron Compton wavelength, $\lambda_\text{C} \sim 1/m(a)$, while GR is recovered on large scales ($k \ll m$).

In this work, we focus on the Hu–Sawicki model~\citep{Hu:2007nk}, which reproduces the \lcdm~background expansion while introducing modifications at the perturbation level. The theory is commonly parametrized by the present-day background value of the scalar field, $f_{R_0}$, which controls the strength and range of the fifth force. For viable values such as $|f_{R_0}| \lesssim 10^{-5}$~\citep{Basilakos:2013nfa, Cataneo:2014kaa}, the model remains consistent with solar system constraints due to the \textit{chameleon screening mechanism}~\citep{Khoury:2003rn}. In high-density environments, the scalaron acquires a large mass, suppressing the fifth force and recovering GR; conversely, in low-density regions like cosmic voids, the force remains active.

The combination of scale-dependent growth, non-linear screening, and environmental dependence makes $f(R)$ gravity a particularly challenging case for fast approximate solvers like \texttt{COLA}, which rely on Lagrangian perturbation theory~\citep{Buchert:1995bq, Matsubara:2007wj, Matsubara:2015ipa} and often lack the force resolution to solve the non-linear field equations accurately. These limitations motivate the development of field-level correction schemes such as \mgnecola. By focusing on this model, we assess the emulator's performance in a regime where both MG signatures and non-linear dynamics are maximally pronounced.

\section{Method}
\label{Sec: Method}

In this section, we describe the simulation datasets employed in this work, including the high-fidelity \mgquijote~suite used as a reference and the approximate simulations generated with \mgcola. While these approximate methods reduce the computational cost by more than three orders of magnitude, their accuracy degrades in the mildly non-linear and fully non-linear regimes—precisely where MG effects are expected to be most prominent. 

Our goal is to develop a methodology that leverages the complementary strengths of both approaches, combining the accuracy of full $N$-body simulations with the efficiency of fast approximate realizations. We then present the CNN architecture adopted in our analysis and describe in detail the training strategy, validation procedure, and performance metrics used to evaluate its predictive accuracy.

\subsection{$N$-Body Simulations}

\subsubsection{Full $N$-Body Simulations: \textnormal{\mgquijote}} 
\label{Sec: N-Body Sims}

As high-fidelity reference data, we employ the \mgquijote~suite~\citep{mgquijote}, a large ensemble of cosmological $N$-body simulations designed to probe extensions of the standard \lcdm~model. Specifically, this suite focuses on the Hu--Sawicki $f(R)$ gravity model. Owing to their fine spatial resolution, these simulations rigorously solve the full non-linear scalar field equations and incorporate the chameleon screening mechanism to yield highly accurate predictions deep into the non-linear regime.

The \mgquijote~suite includes both a set of simulations with fixed cosmological parameters, varying only $f_{R_0}$, and a separate set in which all cosmological and MG parameters are varied following a Sobol sequence that fills a Latin hypercube. For our training, we exclusively utilize the fixed-cosmology set. Fixing the background cosmology forces the CNN to learn only the modifications introduced by the fifth force and the chameleon screening, avoiding possible degeneracies between $f_{R_0}$ and other cosmological parameters. Additionally, this approach avoids the curse of dimensionality inherent in training 3D field-level emulators across multi-parameter spaces, allowing the emulator to converge without requiring an extensive training dataset

The simulations evolve $N_p = 512^3$ CDM particles in a periodic box of side length $L = 1000~h^{-1} \, \Mpc$ from an initial redshift of $z=127$ to $z=0$. This set adopts a fixed cosmology defined by:
\begin{align}
    \Omega_m &= 0.3175, \quad \Omega_b = 0.049, \quad h = 0.6711, \nonumber \\
    n_s &= 0.9624, \quad \sigma_8 = 0.834, \quad M_\nu = 0, \label{Eq: Fiducial cosmology}
\end{align}
where $\Omega_m$ and $\Omega_b$ represent the total matter and baryon density parameters, respectively, $h$ is the reduced Hubble constant, $n_s$ is the primordial scalar spectral index, $\sigma_8$ denotes the root-mean-square amplitude of linear matter fluctuations on scales of $8~h^{-1}\,\Mpc$, and $M_\nu$ is the total neutrino mass. Initial conditions are generated at $z = 127$ using the \texttt{NGenIC} code\footnote{\url{https://github.com/franciscovillaescusa/N-GenIC_growth}}. By employing the \texttt{MG-Gadget} code~\citep{Puchwein:2013lza}, the scalar degree of freedom is evolved dynamically, ensuring a self-consistent modeling of the fifth force and the chameleon screening mechanism. Consequently, while MG effects persist in low-density environments, GR is effectively recovered in high-density regions, such as the interiors of massive collapsed halos.

In addition to the standard cosmological parameters, the suite explores four values of the background scalar field amplitude, $f_{R_0}$, labeled as:
\begin{align}
    &\{f_{R_\text{p}}, f_{R_\text{pp}}, f_{R_\text{ppp}}, f_{R_\text{pppp}}\} = \nonumber \\
    &\left\{ -5\times 10^{-7}, -5\times 10^{-6}, -5\times 10^{-5}, -5\times 10^{-4} \right\}. \label{Eq: HS Params}
\end{align}
For each case, the dataset provides 500 independent realizations, enabling precise statistical characterization of MG effects.

These simulations achieve $\sim 1\%$ force accuracy for modes with $k < 1.6~h\,\Mpc^{-1}$ and reveal distinct MG signatures relative to \lcdm. For instance, they exhibit a $\sim 20\%$ enhancement in matter clustering, quantified by the boost in the MPS, for modes with $k \gtrsim 0.1~h\,\Mpc^{-1}$ for $|f_{R_0}| = 5 \times 10^{-5}$. Beyond two-point statistics, the non-linear dynamics exhibit non-Gaussian features that can be probed using higher-order observables. Phase-matched initial conditions are employed to enable accurate numerical derivatives for Fisher matrix forecasts, while snapshots at $z = \{0, \,0.5, \,1, \,2, \,3\}$ facilitate studies of the redshift-dependent growth of structure~\citep{Valogiannis:2024rvt}.

\subsubsection{Approximate $N$-Body Simulations: \textnormal{\mgcola}}

To efficiently simulate LSS formation in MG models, such as the Hu--Sawicki $f(R)$ theory, the \mgcola~method\footnote{\url{https://github.com/HAWinther/MG-PICOLA-PUBLIC.git}} provides a computationally tractable alternative to full $N$-body solvers like \texttt{MG-Gadget}, on which \mgquijote~is based. The approach extends the COLA framework to MG scenarios by modifying both the particle dynamics and the underlying gravitational potential to account for fifth-force effects and screening mechanisms~\citep{Wright:2017dkw, Winther:2017jof}.

In \mgcola, large-scale modes are evolved using Lagrangian perturbation theory, while the non-linear small-scale dynamics are captured via a Particle-Mesh (PM) solver~\citep{Brandbyge:2008js} with modified equations of motion. For Hu--Sawicki gravity, this requires solving a modified Poisson equation that incorporates the scalaron and its coupling to matter. Screening is implemented through semi-analytical approximations based on the chameleon mechanism, allowing for efficient evaluation without full field solvers. This hybrid approach dramatically reduces computational costs compared to full MG solvers, effectively reducing the computation time by around three order of magnitude. However, this increment in the speed comes with the drawback that accuracy degrades on the smaller scales, exactly where MG effects are most prominent.

In our work, we adopt \mgcola~to perform fast Hu--Sawicki $f(R)$ simulations. Each \mgcola~run is initialized with the same random seeds, cosmological parameters, and simulation parameters (e.g., boxsize, number of particles) as its \mgquijote~counterpart using second order Lagrangian perturbation theory~\citep{Crocce:2006ve, Jenkins:2010} at $z = 9$. Particles are evolved using a PM solver up to $z = 0$. Then, ICs are computed again but at $z = 127$ with no further evolution required, ensuring particle-by-particle correspondence between the two simulations at $z = 127$ and $z = 0$. Although the ICs in \mgquijote~and \mgcola~are computed using different codes, their differences are negligible at such high redshift.

Finally, the dataset consists of 100 paired simulations with the MG parameter fixed following the same labels in Eq.~\eqref{Eq: HS Params}, while the cosmological parameters are held constant as in Eq.~\eqref{Eq: Fiducial cosmology}.

\subsection{Machine Learning Model}

\subsection{Network Architecture, Training and Evaluation}

\begin{figure*}
    \centering
    \includegraphics[width=\linewidth]{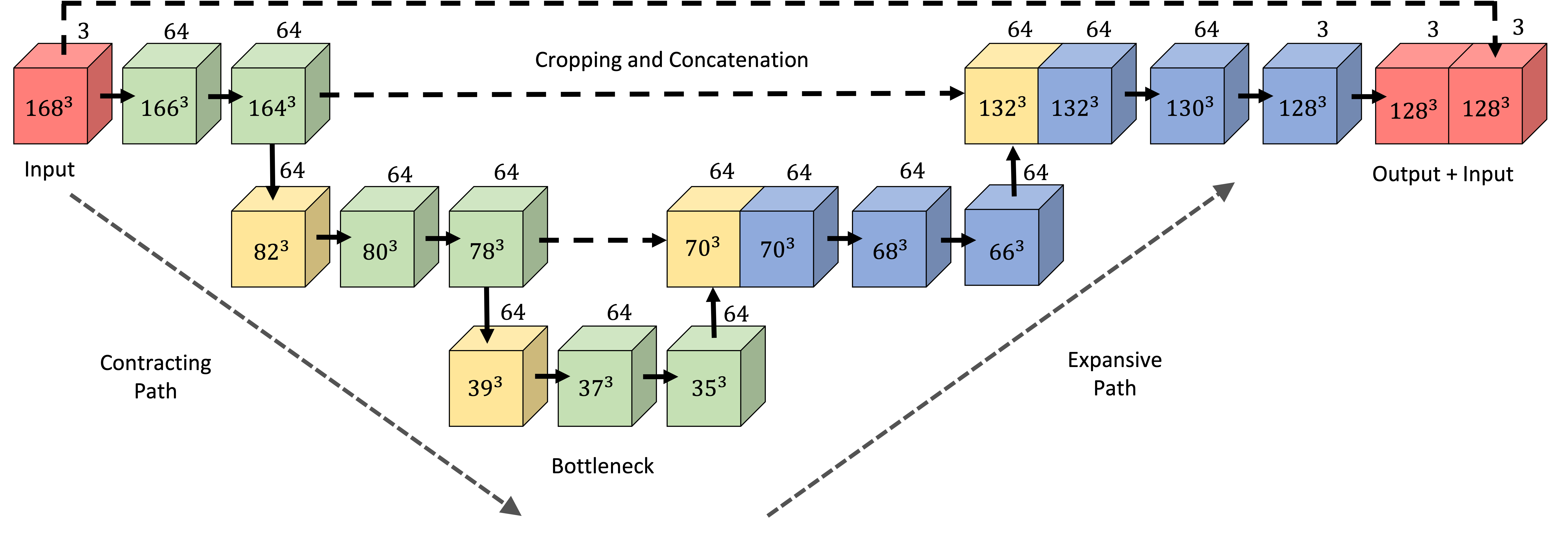}
    \caption{Schematic representation of the V-Net architecture adopted in this work. The network follows an encoder--decoder (contracting--expansive) design with two down-sampling and two up-sampling stages, connected through residual and skip connections at multiple resolutions. The input to the network consists of cubic sub-volumes of size $128^3$ containing the three components of the displacement field (three input channels), which are padded with 20 voxel, resulting in an effective input size of $168^3$. In the first contracting block, two consecutive $3^3$ convolutional layers are applied, reducing the spatial dimensions to $166^3$ and $164^3$, respectively. A residual connection is implemented via a $1^3$ convolution linking the block input to its output. Down-sampling is then performed using a stride-2 $2^3$ convolution, yielding a feature map of size $82^3$. This procedure is repeated to construct the second contracting block, whose output constitutes the bottleneck of the network. The expansive path is initiated by a stride-$1/2$ $2^3$ transposed convolution, increasing the spatial resolution to $70^3$. The resulting feature map is concatenated with the correspondingly cropped output of the matching contracting block, forming a skip connection that preserves small-scale information. Two $3^3$ convolutions are then applied, followed by a residual connection within the block. An analogous sequence defines the second up-sampling stage, ultimately restoring the spatial resolution to $128^3$. The final output of the network is a displacement field of size $128^3$, representing the correction $\mathbf{x}_{f,\mathrm{res}} - \mathbf{x}_{f,\mathrm{MG}\text{-}\mathrm{PICOLA}}$. By adding this correction to the input displacement field, the network predicts the full final displacement $\mathbf{x}_{f,\mathrm{res}} - \mathbf{x}_{i,\mathrm{MG}\text{-}\mathrm{PICOLA}}$. Since ICs are approximately equivalent for both simulations, the resulting field provides a direct approximation to the target high-fidelity displacement, enabling iterative retraining of the model.}
    \label{Fig: VNet}
\end{figure*}

Our network design is based on the convolutional framework introduced in the \texttt{NECOLA} model \citep{Kaushal:2021hqv}, originally developed for \lcdm~simulations. While the underlying architecture is largely preserved, this work introduces two key extensions. First, the framework is generalized to MG cosmologies, where non-linear effects are stronger and exhibit a richer scale dependence. Second, we implement an adapted learning objective that simultaneously enforces accuracy in both Lagrangian and Eulerian displacement fields, complemented by a gradient-matching term that enhances the recovery of small-scale features. These modifications enable \mgnecola~to faithfully capture characteristic MG signatures, going beyond the original \lcdm-only application. The novelty of our approach therefore lies not in the architectural form itself, but in its MG-informed training strategy and its application to non-standard gravity simulations.

Figure~\ref{Fig: VNet} illustrates the V-Net architecture adopted in this work. The network follows a symmetric encoder--decoder (``U-shaped'') V-Net architecture with residual and skip connections at multiple resolutions~\citep{Ronneberger:2015, Milletari:2016arx}. The model has been trained using the \texttt{map2map} code~\citep{AlvesdeOliveira:2020yix}\footnote{\url{https://github.com/eelregit/map2map.git}}. It consists of two down-sampling and two up-sampling stages, where each stage is built from residual blocks containing two unpadded $3^3$ convolutional layers connected through $1^3$ shortcut paths. Down-sampling and up-sampling operations are implemented using $2^3$ convolutions and transposed convolutions with stride 2 and $1/2$, respectively. Since all $3^3$ convolutions are unpadded, encoder feature maps are cropped before concatenation with the corresponding decoder layers. Batch normalization is applied after every convolution except for the first and last two layers, and leaky ReLU activations with slope $0.01$ follow each normalization, as well as the first and penultimate convolutions. All hidden layers employ 64 feature channels, while the input and output layers contain three channels corresponding to the displacement components. Feature maps concatenated through skip connections have 128 channels. 

Due to the large dimensionality of the input displacement fields ($3 \times 512^3$), the data are decomposed into cubic sub-volumes of size $3 \times 128^3$, corresponding to $250~h^{-1}\,\mathrm{Mpc}$ per side. To preserve translational equivariance, no padding is applied to the $3^3$ convolutions. Instead, each sub-cube is periodically padded with 20 voxels per side prior to entering the network, resulting in effective inputs of size $3 \times 164^3$. Data augmentation is performed through random rotations and parity transformations, ensuring equivariance of the displacement field under symmetry operations.

Training is performed using the Adam optimizer~\citep{adam}, with a learning rate of $10^{-4}$, and momenta $\beta_1=0.9$, and $\beta_2=0.999$. We implement a learning rate scheduler to halve the learning rate after three consecutive epochs without improvement in the validation loss. The network is trained for 100 epochs using 25 randomly chosen simulations from each label in Eq.~\eqref{Eq: HS Params}, i.e., a total of 100 simulations, leaving 70\% of them for training, 20\% for validation, and the remaining 10\% for testing. 

Each particle’s trajectory is described by its displacement, defined as $\Delta = \mathbf{x}_f - \mathbf{x}_i$, where $\mathbf{x}_i$ denotes the initial Lagrangian grid positions and $\mathbf{x}_f$ the final positions at the present epoch. The model learns only the residual correction relative to the high-fidelity \mgquijote~reference,
\begin{align}
  \Delta_{\rm res} &= \Delta_\mathrm{QUIJOTE\text{-}MG} - \Delta_\mathrm{MG\text{-}PICOLA} \nonumber \\
   &= \mathbf{x}_{f, \mathrm{QUIJOTE\text{-}MG}} - \mathbf{x}_{f, \mathrm{MG\text{-}PICOLA}},
\end{align}
where we have used the fact that ICs are approximately equivalent in both simulations. Residual learning is implemented by adding the input \mgcola~displacement field to the network output (global bypass connection), such that the corrected displacement field is then obtained as: 
\begin{equation}
    \Delta_\mathrm{MG\text{-}PICOLA} + \Delta_{\rm res} = \mathbf{x}_{f, \mathrm{res}} - \mathbf{x}_{i, \mathrm{MG\text{-}PICOLA}},
\end{equation}
yielding an enhanced approximation that closely reproduces the full $N$-body result at a fraction of the computational cost. 

\subsubsection{Loss Function}

To ensure the accurate reconstruction of the displacement field while simultaneously capturing the non-linear clustering properties of the density field across a wide range of scales, we define a composite loss function. The baseline real-space loss, $\mathcal{L}_{\text{base}}$, is formulated as a weighted sum of three distinct components:
\begin{equation}
\label{Eq: Loss Function}
    \mathcal{L}_\text{base} \equiv w_\mathrm{lag}\mathcal{L}_\mathrm{lag} 
    + w_\mathrm{eul}\mathcal{L}_\mathrm{eul} 
    + w_\mathrm{grad}\mathcal{L}_\mathrm{grad}.
\end{equation}
Here, $\mathcal{L}_{\mathrm{lag}}$ and $\mathcal{L}_{\mathrm{eul}}$ represent the root-mean-squared error (RMSE) of the predicted Lagrangian displacement field and the resulting Eulerian overdensity field, respectively. The third term, $\mathcal{L}_{\mathrm{grad}}$, is a gradient-based penalty designed to regularize the displacement predictions, preventing both unphysical over-smoothing and excessively sharp discontinuities. This gradient term is critical for recovering small-scale structures, which are particularly sensitive to the non-linear dynamics inherent in MG scenarios. Based on validation set performance, we adopt weights of $(w_{\mathrm{lag}}, w_{\mathrm{eul}}, w_{\mathrm{grad}}) = (1.0, 3.0, 2.0)$ to ensure each term contributes equitably to the total gradient during training.

While $\mathcal{L}_{\text{base}}$ effectively constrains real-space features, it does not explicitly enforce the global clustering statistics. To refine the model's performance in the Fourier domain, we incorporate a spectral loss term:
\begin{equation}
\label{Eq: spectral loss}
    \mathcal{L}_\text{spec} = \left\langle w(k) \left[\log P_\text{pred}(k) - \log P_\text{targ}(k) \right]^2 \right\rangle,
\end{equation}
where $P(k)$ denotes the power spectrum of the Eulerian density field and the brackets indicate an average over Fourier modes. To prioritize the recovery of high-wavenumber information, we introduce a scale-dependent weighting function:
\begin{equation}
\label{Eq: weighting function term}
    w(k) = \left(\frac{k^2}{k_\text{Ny}^2}\right) \Theta(k - k_\text{min}),
\end{equation}
which emphasizes small-scale modes above a threshold $k_\text{min}$ relative to the Nyquist wavenumber $k_\text{Ny}$ (see Sec.~\ref{Sec: Performance Metrics}), and excluding large-scale modes that are already well-constrained by the real-space loss terms. We set $k_{\min}=0.3\,h\,\mathrm{Mpc}^{-1}$, which marks the onset of the mildly non-linear regime where approximate methods like \mgcola~begin to deviate from full $N$-body solutions. By restricting the spectral constraint to this range, the loss is focused specifically on the scales where the network must learn corrections to recover the target clustering statistics. The total loss function is thus:
\begin{equation}
\label{Eq: full loss function}
    \mathcal{L} = \mathcal{L}_\text{base} + \mathcal{L}_\text{spec}.
\end{equation}

This multi-objective optimization framework represents a significant evolution from previous approaches, such as \texttt{NECOLA}, which utilized a simplified logarithmic product:
\begin{equation}
\label{Eq: Necola Loss}
    \mathcal{L}_\mathrm{NECOLA} \equiv \log \left(\mathcal{L}_\mathrm{eul}\mathcal{L}_\mathrm{lag}^\lambda \right),
\end{equation}
with a hyperparameter $\lambda$ regulating the trade-off between the Eulerian and Lagrangian losses, since the CNN needs to learn three real numbers for the Lagrangian positions, while the CNN only needs one real number to capture the Eulerian density. Our formulation, by contrast, is specifically tailored to capture the delicate small-scale signatures of MG. The empirical advantage of our composite loss over the \texttt{NECOLA} approach is validated by the ablation study presented in Fig.~\ref{Fig: Ablation}.

\section{Results}
\label{Sec: Results}

In this section, we assess the fidelity and robustness of the \mgnecola\ framework. We benchmark our CNN-enhanced predictions against two baselines: the input approximate simulations (\mgcola) and the ground-truth $N$-body calculations (\mgquijote). Our evaluation relies on a suite of cosmological summary statistics, with a primary focus on the MPS, $P(k)$, and bispectrum, $B(k)$, to quantify accuracy deep into the non-linear regime. We first present an ablation study of our loss function and, then, the statistical performance on the test set. Subsequently, we examine the emulator's generalization capabilities by testing its accuracy on cosmologies with massive neutrinos—a parameter space not encountered during training.

\subsection{Performance Metrics}
\label{Sec: Performance Metrics}

To quantify the statistical agreement between the emulator and the $N$-body ground truth, we utilize the transfer function of the MPS, defined as:
\begin{equation}
    T(k) \equiv  \sqrt{\frac{P_\mathrm{pred}(k)}{P_\mathrm{targ}(k)}},
\end{equation}
where $P_\mathrm{pred}(k)$ and $P_\mathrm{targ}(k)$ denote the auto-power spectra of the predicted field and the target $N$-body simulation, respectively. Values of $T(k) \approx 1$ indicate high fidelity in the recovery of clustering amplitude.

We evaluate our results up to a wavenumber of $k \approx 1.5~h\,\mathrm{Mpc}^{-1}$. We note that this extends slightly beyond the strict trust region of the simulations. The Nyquist frequency of the \mgquijote\ suite is given by:
\begin{equation}
    k_\mathrm{Ny} \equiv \frac{\pi N_p^{1/3}}{L} = 1.61~h\,\mathrm{Mpc}^{-1}.
\end{equation}
While numerical artifacts (aliasing, shot noise) typically limit the reliability of $N$-body spectra to $k \lesssim 0.7 k_\mathrm{Ny} \approx 1.1~h\,\mathrm{Mpc}^{-1}$~\citep{Sefusatti:2015aex}, we include modes up to $k \approx 1.5~h\,\mathrm{Mpc}^{-1}$ to assess the numerical stability of the network at the grid resolution limit.

\subsection{Performance of the New Loss Function}

\begin{figure}[t]
    \centering
    \includegraphics[width=\columnwidth]{Ablation_Acc.pdf}
    \caption{Minimal ablation study comparing the accuracy of our proposed gradient-enhanced loss function (Eq.~\eqref{Eq: Loss Function}, blue) against the standard \texttt{NECOLA} loss (Eq.~\eqref{Eq: Necola Loss}) with varying weights ($\lambda=1, 2, 3$, orange, purple, red, respectively), and the \mgcola~input (green). The proposed loss yields consistently lower errors, particularly on non-linear scales ($k > 0.5~h\,\mathrm{Mpc}^{-1}$), keeping the accuracy below $1\%$ on all scales. The $1\%$ error region is bounded by a horizontal black dotted line, and the scale $k = 1~h\,\Mpc^{-1}$ is marked with a vertical black dotted line.}
    \label{Fig: Ablation}
\end{figure}

To quantify the impact of the gradient-matching term ($\mathcal{L}_\mathrm{grad}$), we performed an ablation study comparing our total loss formulation against the original \texttt{NECOLA} loss function (defined in Eq.~\eqref{Eq: Necola Loss}), and the \mgcola~input. Figure~\ref{Fig: Ablation} displays the percentage deviation of the transfer function from unity, $|1 - T(k)|\times100$, for a representative realization from the test set.

It is evident that removing the gradient term and reverting to the baseline \texttt{NECOLA} loss significantly degrades performance. The degradation is most pronounced in the non-linear regime ($k > 0.5~h\,\mathrm{Mpc}^{-1}$), where the baseline models (shown for $\lambda \in \{1, 2, 3\}$) exhibit errors up to 2.5\% larger than our method. This confirms that explicitly penalizing discrepancies in the displacement gradients forces the network to better resolve the small-scale clustering features, which are essential for capturing MG signatures.

\subsection{Distribution of Residuals}

\begin{figure}[t]
    \centering
    \includegraphics[width=\columnwidth]{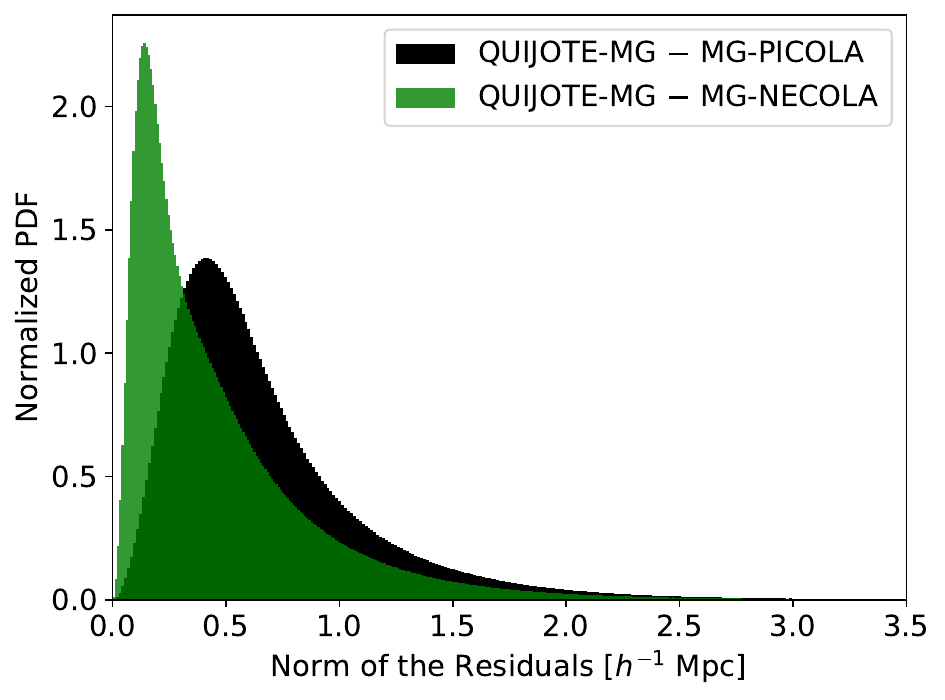}
    \caption{Probability density function of the position error magnitude (residuals) for the approximate \mgcola\ simulation (black) and the emulator \mgnecola\ (green) relative to the \mgquijote\ $N$-body ground truth. The \mgnecola\ distribution is sharply peaked near zero, indicating a substantial improvement in particle-level positional accuracy.}
    \label{Fig: Residuals}
\end{figure}

\begin{figure}[t]
    \centering
    \includegraphics[width=\columnwidth]{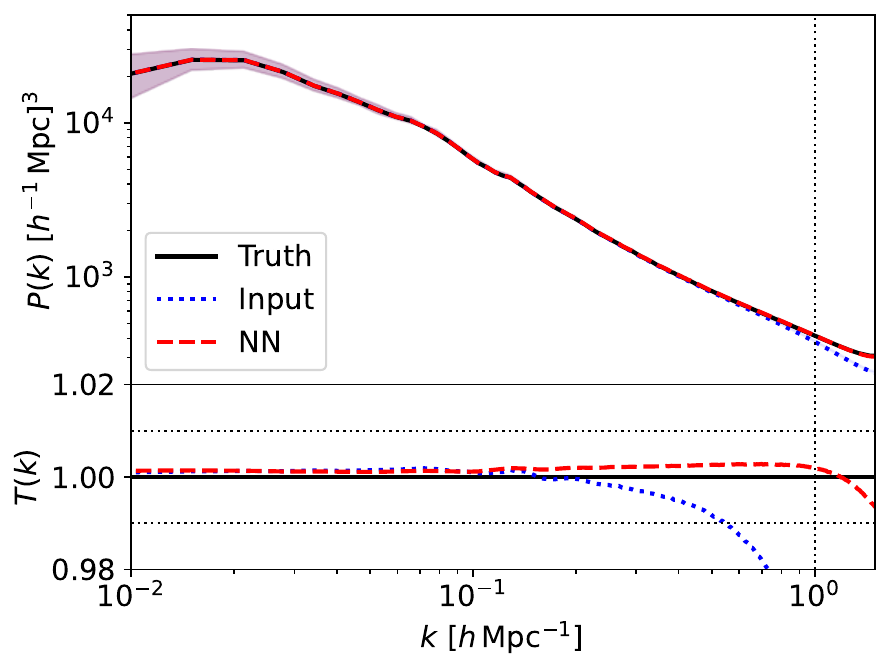}
    \caption{The monopole of the MPS, $P(k)$, with its $1\sigma$ region (top), and the transfer function, $T(k)$ (bottom). The approximate \mgcola~input (blue dotted) deviates significantly at small scales, while \mgnecola~(red dashed) recovers the \mgquijote~$N$-body truth (black solid) with sub-percent precision.}
    \label{Fig: Pk}
\end{figure}

\begin{figure}[t]
    \centering
    \includegraphics[width=\columnwidth]{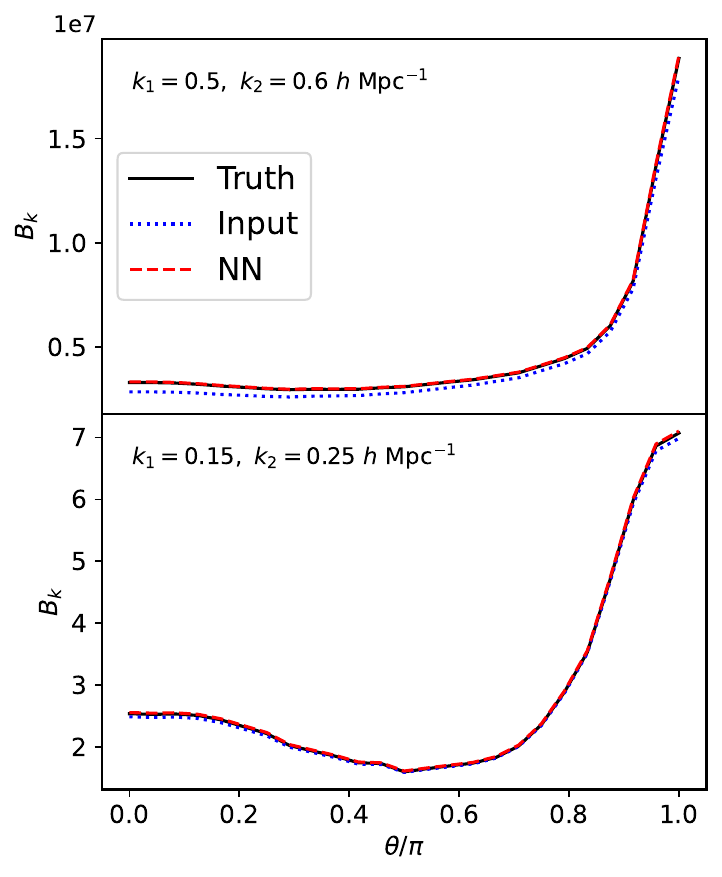}
    \caption{The bispectrum $B(k_1, k_2, \theta)$ for two triangle configuration: (top) $k_1=0.5$ and $k_2=0.6~h\,\mathrm{Mpc}^{-1}$, (bottom) $k_1=0.15$ and $k_2=0.25~h\,\mathrm{Mpc}^{-1}$. Top panel shows that \mgnecola~correctly enhances the \mgcola~output when mildly non-linear scales are considered.}
    \label{Fig: Bk}
\end{figure}

While the power spectrum and bispectrum quantify clustering statistics in Fourier space, they do not directly assess the spatial precision of the displacement field at the field level. To verify that \mgnecola\ improves the actual trajectories of dark matter particles, we examine the distribution of the displacement residuals for a particular realization in the test set. We define the residual $\delta r$ for each particle as the Euclidean norm of the vector difference between the predicted displacement and the $N$-body truth:
\begin{equation}
    \delta r \equiv |\mathbf{d}_\mathrm{pred} - \mathbf{d}_\mathrm{truth}|.
\end{equation}

Figure~\ref{Fig: Residuals} presents the probability density function (PDF) of these residuals. The input \mgcola~simulations (black) exhibit a broad distribution with a mean residual error of $\langle \delta r \rangle \approx 0.68~h^{-1}\,\mathrm{Mpc}$ and a heavy tail extending beyond $2.0~h^{-1}\mathrm{Mpc}$. This reflects the inherent approximations of the 2LPT solver, which accumulate errors particularly in the non-linear regime.

In contrast, the \mgnecola~predictions (green) show a dramatic shift in the distribution. The PDF is sharply peaked below $0.5~h^{-1}\mathrm{Mpc}$, effectively suppressing the high-error tail present in the input. Quantitatively, the neural network reduces the mean displacement error to $\langle \delta r \rangle \approx 0.47~h^{-1}\mathrm{Mpc}$, representing an improvement of approximately $30\%$ in particle positioning accuracy. This tightening of the distribution demonstrates that the neural network does not merely act as a diffusive filter but actively corrects the displacement vectors particle-by-particle, aligning the resulting dark matter sheet more closely with the full $N$-body solution.

\subsection{Power Spectrum and Bispectrum}

We now turn to the statistical recovery of the matter field. Figure~\ref{Fig: Pk} presents the performance on the test set of \mgnecola~(red dashed lines) compared to the approximate \mgcola~input (blue dotted lines) and the ground-truth \mgquijote\ $N$-body simulations (black solid lines). The $1\%$ error region is bounded by two horizontal black dotted lines, and the scale $k = 1~h\,\Mpc^{-1}$ is marked with a vertical black dotted line.

This figure displays the MPS, $P(k)$, and the associated transfer function, $T(k)$. The shaded regions denote the $1\sigma$ standard deviation computed across the 10 independent test realizations. We observe a pronounced variance at large scales ($k \lesssim 0.1~h\,\mathrm{Mpc}^{-1}$), which corresponds to the expected cosmic variance arising from the different ICs (random seeds) used in each realization.

While the approximate solver (\mgcola) begins to lose power significantly at $k > 0.5~h\,\mathrm{Mpc}^{-1}$—underestimating the clustering amplitude by more than 5\% at non-linear scales—the \mgnecola~emulator successfully restores the power. The transfer function demonstrates that our method recovers the $N$-body amplitude to within $1\%$ accuracy up to $k = 1.0~h\,\mathrm{Mpc}^{-1}$ and beyond, effectively correcting the deficit inherent to the COLA method.

Figure~\ref{Fig: Bk} assesses the non-Gaussian information content via the bispectrum, $B(k_1, k_2, \theta)$. We show two representative triangle configurations with fixed sides: top panel shows $k_1=0.5~h\,\mathrm{Mpc}^{-1}$ and $k_2=0.6~h\,\mathrm{Mpc}^{-1}$, while bottom panel shows $k_1=0.15~h\,\mathrm{Mpc}^{-1}$ and $k_2=0.25~h\,\mathrm{Mpc}^{-1}$ as a function of the opening angle $\theta$. Here, the improvement is equally pronounced; \mgnecola~corrects the shape-dependent bias of the input simulation, tracking the $N$-body bispectrum shape with high fidelity across all angular configurations when mildly non-linear scales are involved.

Collectively, these results confirm that \mgnecola~learns the mapping from approximate to fully non-linear gravity, capturing both the 2-point clustering amplitude and the mode-coupling signatures encoded in the 3-point function.

\subsubsection{Computational Efficiency}

\begin{figure*}[t]
    \centering
    \begin{minipage}{0.45\textwidth}
        \centering
        \includegraphics[width=\linewidth]{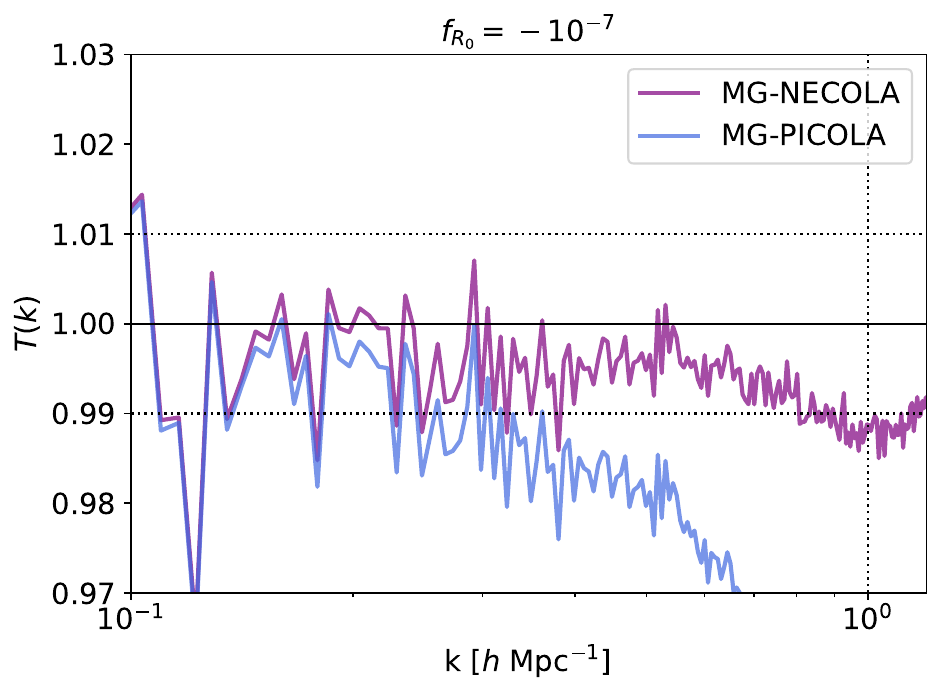}
    \end{minipage}
    \hfill
    \begin{minipage}{0.45\textwidth}
        \centering
        \includegraphics[width=\linewidth]{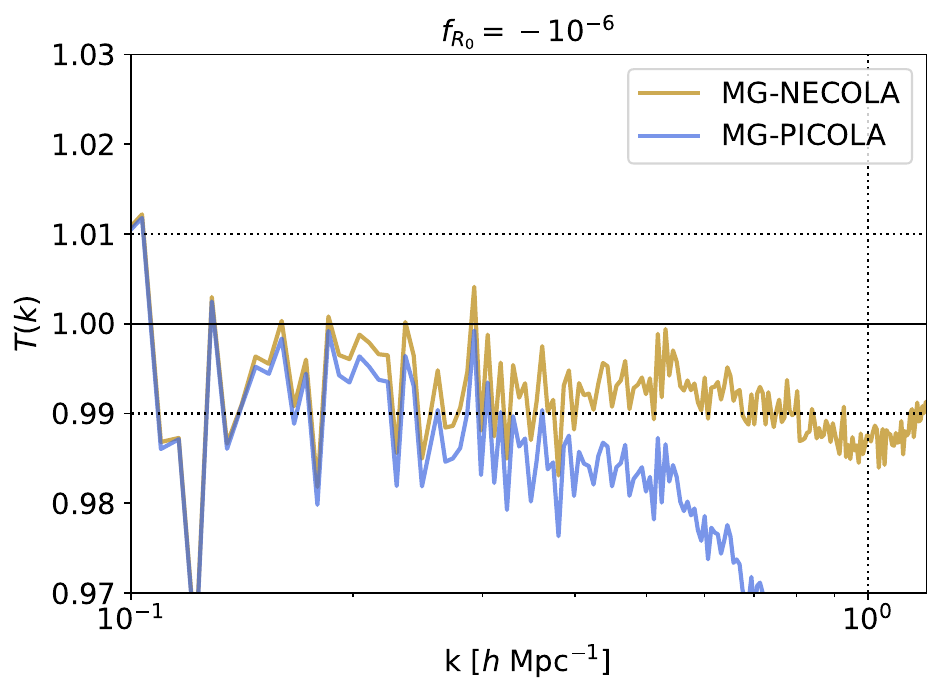}
    \end{minipage}
    
    \vspace{0.2cm} 
    
    \begin{minipage}{0.45\textwidth}
        \centering
        \includegraphics[width=\linewidth]{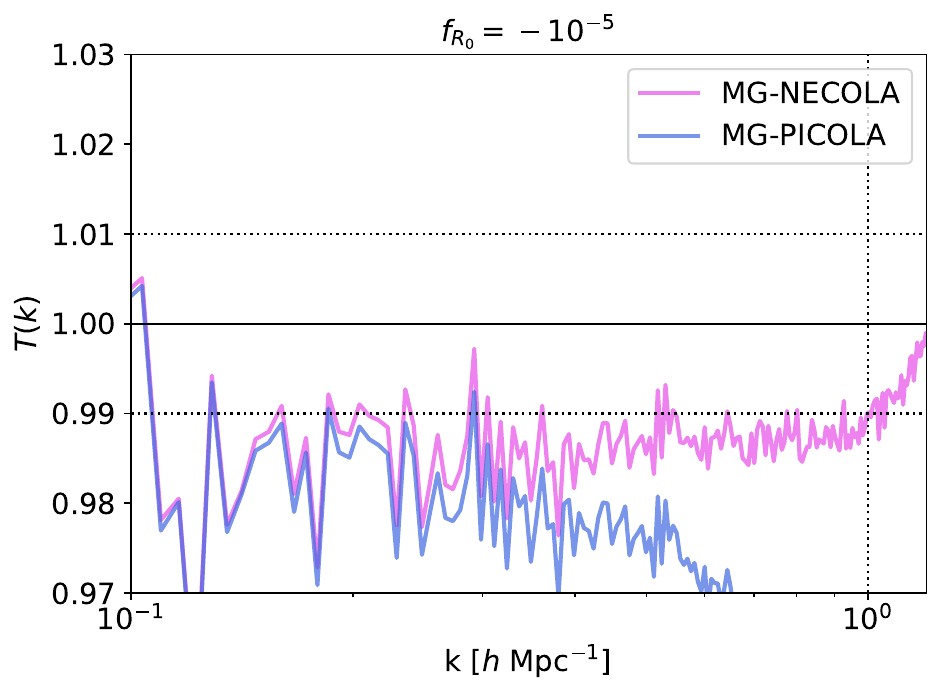}
    \end{minipage}
    \hfill
    \begin{minipage}{0.45\textwidth}
        \centering
        \includegraphics[width=\linewidth]{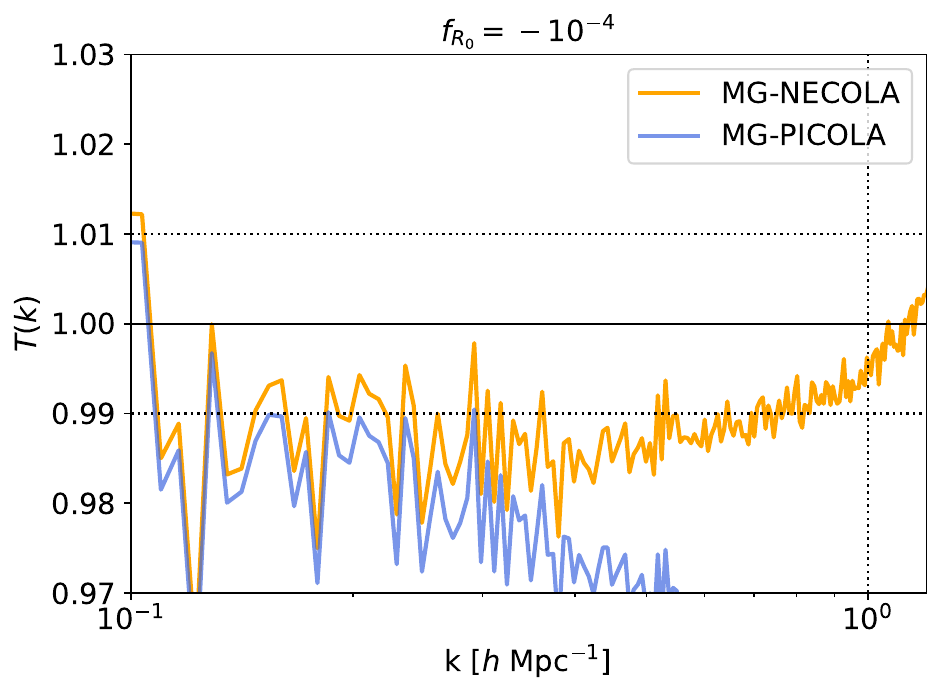}
    \end{minipage}
    \caption{Validation against the \texttt{fRemu} power spectrum emulator across varying scalar field strengths. We fix the cosmology to fiducial and vary $|f_{R_0}|$ from $10^{-7}$ (top-left) to $10^{-4}$ (bottom-right). The plots show the transfer function relative to the \texttt{fRemu} prediction. Large scales ($k < 0.1~h\,\mathrm{Mpc}^{-1}$) are excluded to avoid comparing the smooth emulator prediction with the cosmic variance of a single realization. In all regimes, \mgnecola~significantly improves upon the approximate solver \mgcola~(lightblue), maintaining agreement with the MPS emulator to within $\sim 1\%$ on non-linear scales.}
    \label{Fig: fRemu}
\end{figure*}

One of the primary motivations for developing \mgnecola~is to bypass the prohibitive computational cost of high-resolution $N$-body simulations for parameter inference pipelines. In Table~\ref{Tab: Timing}, we summarize the computational resources required for each method.

A full $N$-body simulation using the \mgquijote~setup requires approximately 500 CPU-hours ($\sim 1.8 \times 10^6$ CPU-seconds) to reach $z=0$. In contrast, the approximate \mgcola~solver drastically reduces this burden to $\sim 10^3$ CPU-seconds, albeit at the cost of accuracy on small scales.

The \mgnecola~emulator acts as a post-processing refinement step. Once trained, the inference stage takes approximately 180 seconds per realization on a single GPU. Consequently, the total time-to-solution for our method remains roughly three orders of magnitude faster than the $N$-body baseline~(see Tab.~\ref{Tab: Timing}). This massive speed-up enables the generation of thousands of high-fidelity mock catalogs required for covariance matrix estimation and likelihood analyses in modified gravity cosmologies.

\begin{table}[t]
    \centering
    \caption{Comparison of computational costs per realization. The ``Total Speed-up'' is calculated relative to the \mgquijote~$N$-body baseline. The time for \mgnecola~includes both the generation of the input COLA simulation and the neural network inference time.}
    \label{Tab: Timing}
    \begin{tabular}{l c c c}
    \hline
    \hline
    Method & Hardware & Time [s] & Speed-up \\
    \hline\hline
    \mgquijote & CPU & $\sim 1.8 \times 10^6$ & $1\times$ \\
    \mgcola & CPU & $\sim 1.0 \times 10^3$ & $1800\times$ \\
    \textbf{\mgnecola} & \textbf{CPU + GPU} & $\mathbf{\sim 1.2 \times 10^3}$ & $\mathbf{\sim 1500\times}$ \\
    \hline\hline
    \end{tabular}
\end{table}

\subsection{\textnormal{\mgnecola}~Extrapolation}

A central challenge for neural emulators is demonstrating their ability to generalize beyond the specific parameter space covered during training. To verify that \mgnecola~has learned a robust, physical mapping of gravitational structure formation—rather than merely memorizing the training manifold—we subject the network to four stringent extrapolation and generalization tests on unseen cosmologies. Specifically, we evaluate the model in regimes that involve distinct physical signatures and configurations: (1) the MPS for MG parameters not considered in the training set while fixing the baseline cosmology, (2) simultaneous variation of all cosmological and MG parameters across a broad hypercube, (3) the GR limit (\lcdm), and the inclusion of massive neutrinos on top of \lcdm~(4) and on top of MG cosmologies (5).

\subsubsection{Sensitivity to Scalar Field Strength}

\begin{figure}[t]
    \centering
    \includegraphics[width=\columnwidth]{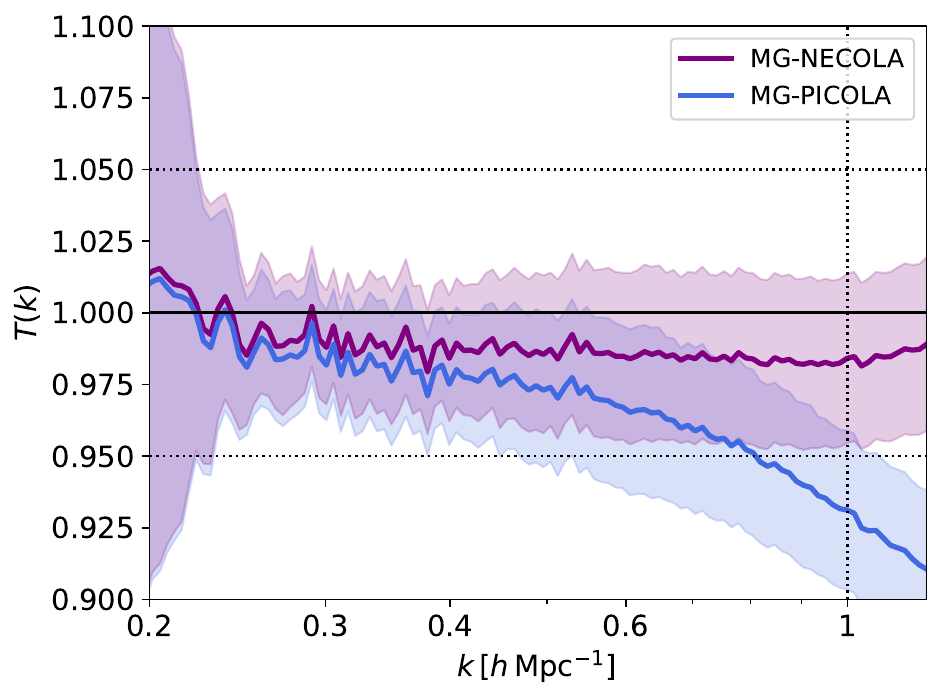}
    \caption{Mean transfer function $T(k)$ and $1\sigma$ dispersion across 98 distinct cosmologies spanning the $\{\Omega_m, \Omega_b, h, n_s, \sigma_8, f_{R_0}\}$ parameter space. While \mgcola~(blue) exhibits a systematic lack of power, \mgnecola~(purple) successfully rectifies the power spectrum. It achieves near-unity accuracy with tight variance (although similar to that of \mgcola~in this respect) up to $k \sim 1.0~h\,\mathrm{Mpc}^{-1}$, demonstrating robust generalization.}
    \label{Fig: Extrapolation}
\end{figure}

\begin{figure*}[t]
    \centering
    \begin{minipage}{0.45\textwidth}
        \centering
        \includegraphics[width=\linewidth]{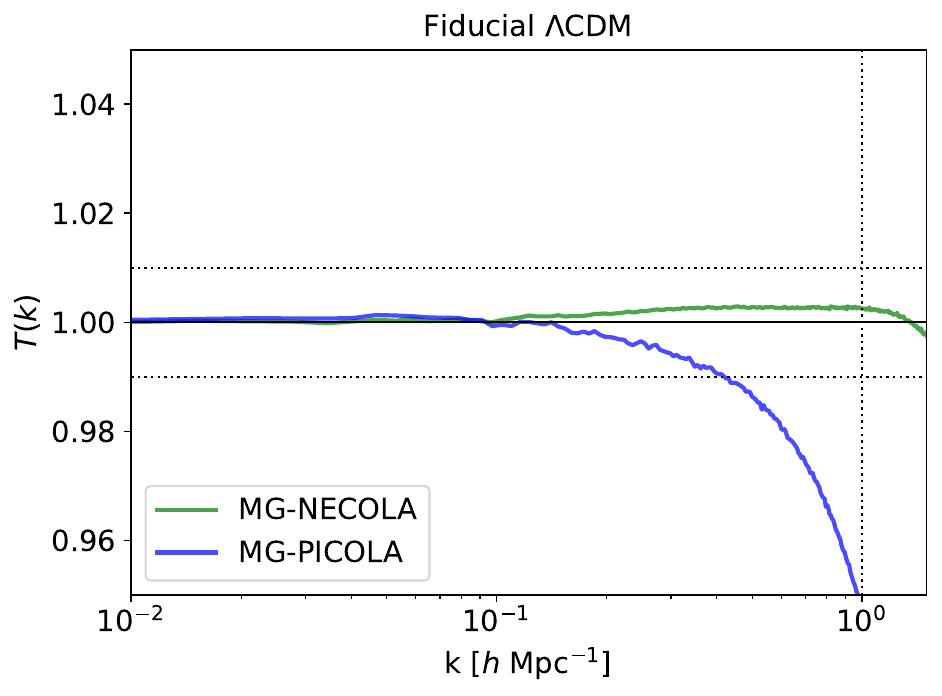}
    \end{minipage}
    \hfill
    \begin{minipage}{0.45\textwidth}
        \centering
        \includegraphics[width=\linewidth]{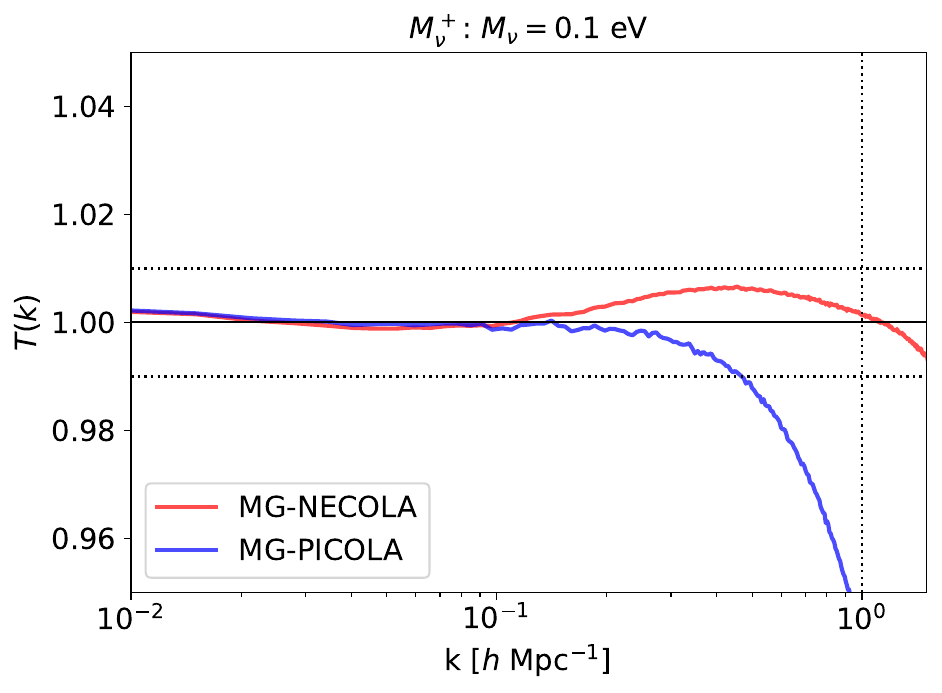}
    \end{minipage}
    
    \vspace{0.2cm} 
    
    \begin{minipage}{0.45\textwidth}
        \centering
        \includegraphics[width=\linewidth]{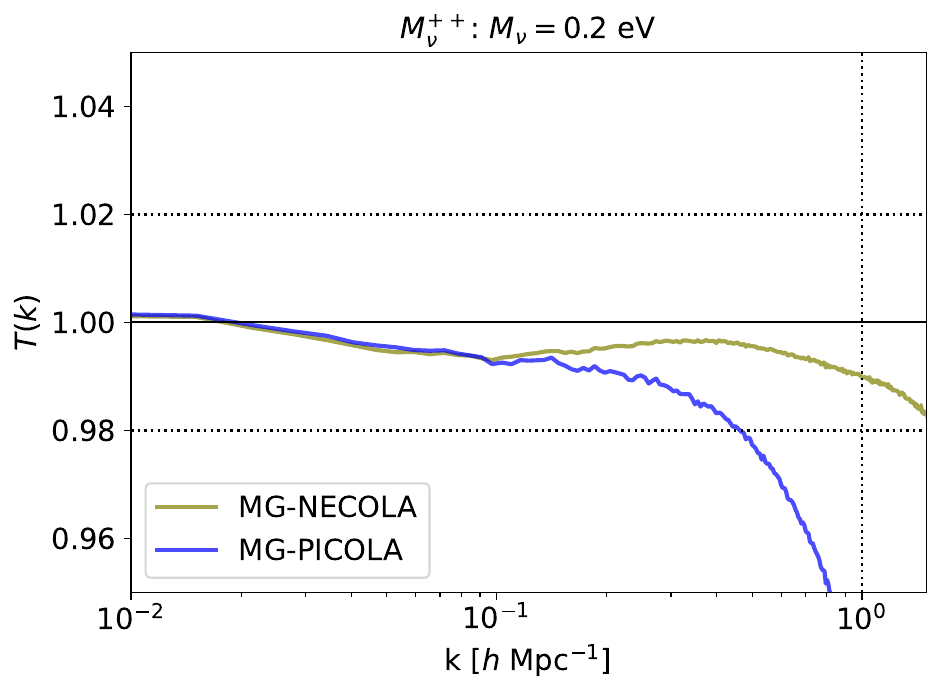}
    \end{minipage}
    \hfill
    \begin{minipage}{0.45\textwidth}
        \centering
        \includegraphics[width=\linewidth]{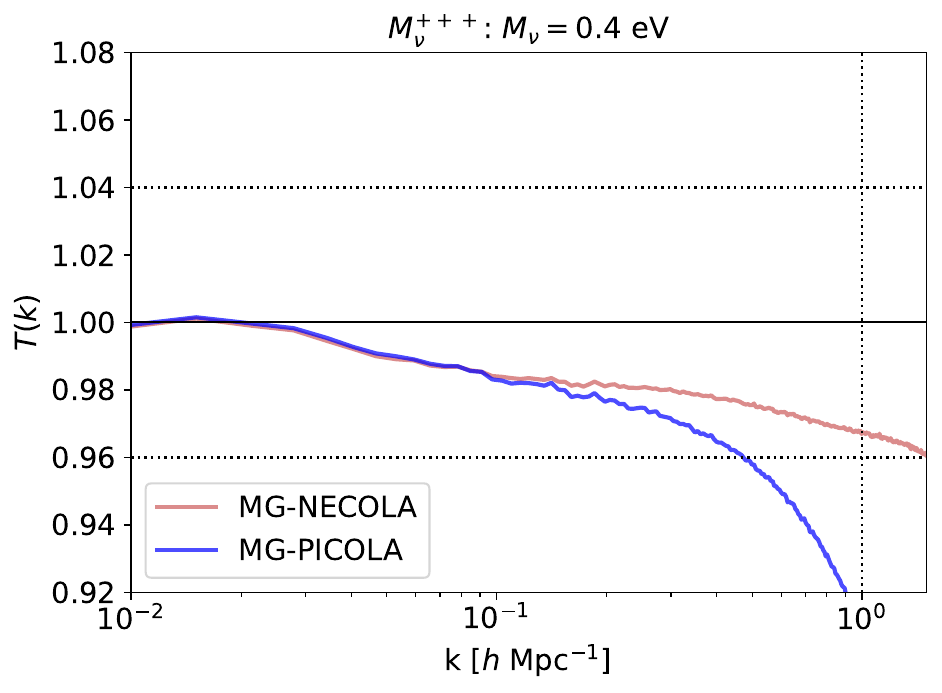}
    \end{minipage}
    \caption{Generalization performance on unseen cosmologies, showing the transfer function $T(k)$ relative to the high-fidelity $N$-body truth. \textbf{Top-Left:} The \lcdm~limit. \textbf{Top-Right to Bottom-Right:} Massive neutrino cosmologies with $M_\nu = 0.1, 0.2, 0.4$ eV, respectively. In all panels, the input approximate solver (blue) deviates significantly at small scales. In contrast, \mgnecola~recovers the $N$-body truth with high fidelity: less than $1\%$ for \lcdm~and $M_\nu^+$, less than 2\% for $M_\nu^{++}$, and less than 4\% for $M_\nu^{+++}$.}
    \label{Fig: Generalization}
\end{figure*}

First, we validate the ability of \mgnecola~to interpolate across the MG parameter space. Here, we compare our predictions against \texttt{fRemu}~\citep{Bai:2024cgt}, a high-precision MPS emulator trained on the full \mgquijote~suite. In a seven dimension space parameter ($\Omega_m$, $\Omega_b$, $h$, $n_s$, $\sigma_8$, $M_\nu$, and $f_{R_0}$), this emulator achieves a better than $95\%$ accuracy. This allows us to sweep through the scalar field amplitude $f_{R_0}$ while keeping the background cosmological parameters fixed to their fiducial values in Eq.~\eqref{Eq: Fiducial cosmology}, to explore the ability of \mgnecola~to capture MG related effects.

We generate \mgnecola~realizations for four distinct field strengths covering the range $|f_{R_0}| \in [10^{-7}, 10^{-4}]$. The results are compared to the \texttt{fRemu} predictions via the transfer function. Figure~\ref{Fig: fRemu} displays the results. We restrict the comparison to $k > 0.1~h\,\mathrm{Mpc}^{-1}$ because \texttt{fRemu} predicts the \textit{mean} power spectrum (smooth), whereas our output comes from a single realization dominated by cosmic variance at large scales; comparing them directly at low $k$ would be dominated by sample noise rather than model performance.

Across all four orders of magnitude of $|f_{R_0}|$, \mgnecola~demonstrates remarkable consistency, tracking the emulator's prediction with $\sim 1\%$ accuracy up to $k=1.0~h\,\mathrm{Mpc}^{-1}$. This test confirms that \mgnecola~has not simply learned a fixed correction kernel, but is sensitive to the specific value of $f_{R_0}$, accurately modulating the displacement corrections to match the theoretical expectation of the Hu-Sawicki model.

\subsubsection{Generalization Across the Cosmological Parameter Space}

While fixing the baseline cosmology provides a clean test of the scalar field dynamics, a truly robust emulator must be able to handle the complex, non-linear degeneracies between gravity and background expansion. To validate this, we evaluate \mgnecola's performance when simultaneously varying the full set of background cosmological parameters alongside the MG scalar field strength.

We generate a Latin hypercube set considering 98 unique cosmologies, spanning a comprehensive parameter space: $\Omega_m \in [0.25, 0.35]$, $\Omega_b \in [0.045, 0.055]$, $h \in [0.62, 0.73]$, $n_s \in [0.92, 1.0]$, $\sigma_8 \in [0.8, 0.86]$, and a broad range of scalar field amplitudes $|f_{R_0}| \in [-10^{-7}, -10^{-3}]$. For each cosmology, we apply the \mgnecola~correction to the approximate \mgcola~output and compare the resulting power spectrum against the high-precision predictions of the \texttt{fRemu} emulator evaluated at the exact same cosmological point.

The results are presented in Figure~\ref{Fig: Extrapolation}, which displays the mean transfer function and its $1\sigma$ dispersion across all 98 cosmologies. The approximate \mgcola~solver (blue) exhibits a large systematic lack of power at small scales, accompanied by a significant scatter, which reflects its varying degree of inaccuracy depending on the specific cosmological background. In stark contrast, \mgnecola~(purple) successfully rectifies the displacement fields across the entire hypercube. The mean transfer function remains centered near unity up to $k=1.0~h\,\mathrm{Mpc}^{-1}$, and the significantly narrower $1\sigma$ band demonstrates that the network's accuracy is stable and reliable regardless of the underlying cosmology. This confirms that \mgnecola~effectively functions as a generalized cosmological emulator, rather than being confined to narrow trajectories in parameter space.

\subsubsection{$\Lambda$\textnormal{CDM}}

We evaluate the \mgnecola~performance in the GR limit (\lcdm), recovered when $|f_{R_0}| = 0$. This represents a parameter point strictly on the boundary of our training prior.

The top-left panel of Figure~\ref{Fig: Generalization} displays the transfer function for the fiducial $\Lambda$CDM cosmology. While the input \mgcola~solver (green) underestimates the power by $\sim 5\%$ at $k=1.0~h\,\mathrm{Mpc}^{-1}$, \mgnecola~(blue) accurately identifies the absence of scalar forces. Instead of introducing spurious MG signals, the network acts purely as a resolution corrector, recovering the $N$-body truth with sub-percent precision ($|1-T(k)| < 0.5\%$). This confirms that \mgnecola~effectively generalizes the original \texttt{NECOLA} framework, encompassing the standard GR scenario as a reliable limit within its expanded parameter space.

\subsubsection{$\Lambda$\textnormal{CDM} and Massive Neutrinos}

A more stringent test involves massive neutrinos, which suppress structure growth on small scales via free-streaming~\citep{Agarwal:2011}—an effect phenomenologically opposite to the enhancement induced by typical $f(R)$ models. Since \mgnecola~was trained exclusively on MG simulations (which boost power), a key question is whether the network acts as a generic ``power booster'' or if it correctly respects the suppression signatures of massive neutrinos.

Here, we utilize the $M_\nu^+$, $M_\nu^{++}$, and $M_\nu^{+++}$ sets from the standard \texttt{QUIJOTE} suite~\citep{Villaescusa-Navarro:2019bje}. These high-resolution $N$-body simulations evolve $512^3$ CDM particles alongside $512^3$ neutrino particles, assuming standard gravity. Each set adopts a fixed cosmology with a total neutrino mass of $M_\nu = 0.1, 0.2,$ and $0.4$~eV, respectively.

The results are shown in the remaining three panels of Figure~\ref{Fig: Generalization}. Remarkably, the emulator greatly improves accuracy across all neutrino masses up to $k \sim 1.0~h\,\mathrm{Mpc}^{-1}$. Even in the extreme case of $M_\nu = 0.4$ eV (bottom-right), where the input approximate solver exhibits significant scale-dependent errors, \mgnecola~correctly identifies the local dynamical state up to $\sim3\%$. It successfully disentangles the resolution corrections from the underlying cosmological model, preserving the physical suppression of the power spectrum. This implies that the network has learned to correct displacement fields based on local density configurations rather than simply memorizing a global amplification factor. 

\subsubsection{$f(R)$ and Massive Neutrinos}

\begin{figure}[t]
    \centering
    \includegraphics[width=\columnwidth]{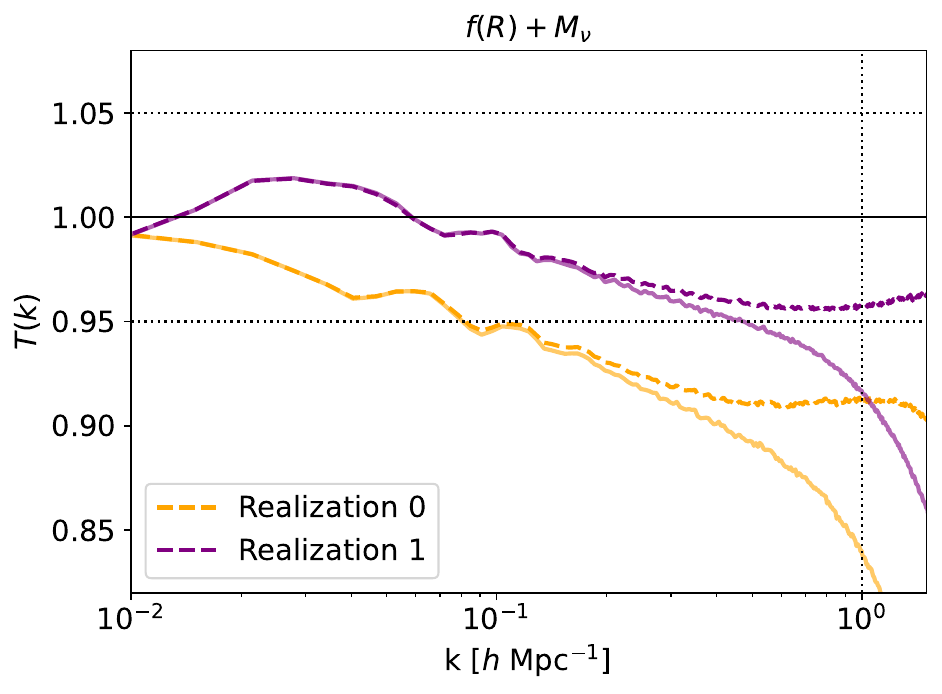}
    \caption{Transfer function $T(k)$ for two independent realizations of the combined scenario ($f(R) + \nu$). The approximate solver \mgcola~(solid lines) exhibits a severe theoretical breakdown in this cases, generating displacement fields that rapidly lose power even at mildly non-linear scales. Due to this heavily degraded baseline, the residual corrector \mgnecola~(dashed lines) cannot fully reconstruct the true $N$-body power spectrum. However, the neural network still attempts to enforce the correct physical clustering, successfully recovering a significant fraction of the missing power. This illustrates both the limitations of the input solver and the robustness of the network's learned physical mapping.}
    \label{Fig: fR_nu}
\end{figure}

Finally, we explore the combined scenario featuring both modified gravity and massive neutrinos. In this regime, the physical effects are highly non-trivial, as the scale-dependent enhancement of structure formation driven by the scalar field competes directly with the free-streaming suppression induced by massive neutrinos~\citep{Baldi:2013iza, Giocoli:2018gqh}.

As previously mentioned, the \mgquijote~suite includes a separate Latin Hypercube set where all cosmological and MG parameters are varied simultaneously. To evaluate the performance of our emulator under these complex conditions, we select two representative samples from this set, whose parameters are summarized in Table~\ref{Tab: LH_Params}.
\begin{table}[t]
\centering
\caption{Cosmological and MG parameters for the two selected realizations from the Latin Hypercube set.}
\label{Tab: LH_Params}
\begin{tabular}{lcc}
\hline\hline
Parameter & Realization 0 & Realization 1 \\ \hline\hline
$\Omega_m$                  & 0.162    & 0.316    \\
$\Omega_b$                  & 0.053    & 0.033    \\
$h$                         & 0.743    & 0.650    \\
$n_s$                       & 0.897    & 1.006    \\
$\sigma_8$                  & 1.109    & 0.782    \\
$M_\nu$ [eV]                & 0.269    & 0.615    \\
$f_{R_0} \times 10^{-4}$     & $-2.026$ & $-0.542$ \\ \hline\hline
\end{tabular}
\end{table}
The full set of cosmological parameters for the \mgquijote~suite is publicly available at the official \href{https://github.com/franciscovillaescusa/Quijote-simulations/blob/master/modified_gravity/Cosmological_parameters.txt}{repository}. 

Figure~\ref{Fig: fR_nu} shows the transfer functions for these independent realizations. Our analysis reveals a significant limitation originating from the baseline approximate solver. Specifically, we find that \mgcola~fails to generate reliable displacement fields in these cases, resulting in transfer functions that diverge rapidly even at mildly non-linear scales.

Because \mgnecola~operates as a residual corrector—mapping approximate local densities to high-fidelity non-linear displacements—its performance is intrinsically tied to the accuracy of the input simulation. When the underlying dynamics provided by the COLA solver are fundamentally compromised, the required adjustment exceeds the scope of a localized correction. Consequently, the network cannot fully reconstruct the $N$-body power spectrum from such degraded initial trajectories.

Nevertheless, the neural network consistently attempts to impose the correct physical behavior. As shown in the plotted samples, \mgnecola~successfully rectifies a substantial portion of the missing power, shifting the transfer functions toward unity (e.g., reaching $T(k) \sim 0.96$ at intermediate scales in Realization 1). This suggests that while the emulator cannot entirely compensate for a failing input solver, the learned mapping remains robust in its tendency to drive the density field toward the correct structural state.

Previous studies, such as \citet{Wright:2017dkw}, indicate that CDM power spectra in combined $f(R)+M_\nu$ scenarios are typically reliable at the percent level up to $k \sim 0.5$--$0.7~h\,\Mpc^{-1}$. However, as seen in Fig.~\ref{Fig: Generalization}, the accuracy of \mgcola~degrades by more than 2\% for $k \gtrsim 0.3~h\,\Mpc^{-1}$ in these cases. We attribute this reduced performance primarily to the high values of $M_\nu$ in these specific realizations. While an exhaustive benchmark between \mgcola~and full $N$-body simulations is beyond the scope of this work, our results highlight the necessity of a sufficiently accurate baseline for field-level emulators.

\section{Summary and Conclusions}
\label{Sec: Conclusions}

In this work, we introduced \mgnecola, a deep learning framework designed to bridge the gap between fast approximate simulations and high-fidelity $N$-body codes in the context of Modified Gravity (MG) cosmologies. By enhancing the output of the computationally efficient \mgcola\ solver with a V-Net architecture, we have developed a pipeline capable of generating accurate mock catalogs for $f(R)$ gravity at a fraction of the computational cost of standard $N$-body methods.

Our analysis demonstrates that \mgnecola\ successfully recovers the non-linear structure of the cosmic web, reducing the mean particle displacement error by approximately $30\%$ relative to the input \mgcola~simulations. This field-level improvement translates into high statistical precision; the emulator recovers the matter power spectrum, $P(k)$, and bispectrum, $B(k)$, of the target \mgquijote\ $N$-body simulations with sub-percent accuracy up to $k \sim 1.0~h\,\mathrm{Mpc}^{-1}$. We find that this performance is largely driven by the inclusion of a gradient-matching term in the loss function, which forces the network to resolve the small-scale displacement gradients responsible for the ``fifth force'' enhancement.

Crucially, we verified that the emulator learns a robust physical mapping rather than merely memorizing the training data. \mgnecola~exhibits remarkable generalization capabilities, correctly interpolating across intermediate scalar field strengths ($|f_{R_0}| \in [10^{-7}, 10^{-4}]$), tracking theoretical predictions consistently. Most notably, we demonstrated the emulator's stability across a broad six-dimensional parameter space ($\{\Omega_m, \Omega_b, h, n_s, \sigma_8, f_{R_0}\}$). Evaluated on a Latin hypercube set consisting of 98 distinct cosmologies, the network successfully rectified the cosmology-dependent systematic errors of the approximate solver, maintaining near-unity accuracy and a tight $1\sigma$ variance up to highly non-linear scales. Furthermore, the network cleanly extrapolates to the General Relativity limit (\lcdm) without introducing spurious MG signals, and accurately captures the power suppression induced by massive neutrinos ($M_\nu \in [0.1, 0.4]$ eV) despite being trained exclusively on power-enhancing MG models.

However, we also identified the fundamental limitations of our residual correction approach when dealing with highly complex, coupled physical regimes. In the combined scenario of modified gravity and massive neutrinos, the baseline \mgcola~solver experiences a severe theoretical breakdown, producing heavily degraded displacement fields in the cases considered here. Because \mgnecola~relies on a reasonably accurate baseline, it cannot fully reconstruct the true $N$-body power spectrum from these fundamentally flawed initial trajectories. Nevertheless, the network still robustly attempts to enforce the correct physical clustering, successfully recovering a significant fraction of the missing power. This demonstrates that while the emulator cannot completely override a failing input solver, its learned physical mapping remains sound.

From a computational perspective, the total time-to-solution for our pipeline is approximately $1500\times$ faster than a full $N$-body run. This efficiency enables the generation of the massive ensembles of mock catalogs required for covariance matrix estimation in future stage-IV survey analyses.

\begin{acknowledgments}
The authors thank Francisco Villaescusa-Navarro for insightful discussions. J.B.O.Q. expresses his gratitude to the Department of Physics at Michigan Technological University for their hospitality during the development of this work. J.B.O.Q., M.R., and E.G. gratefully acknowledge the IT Department at Michigan Technological University for their technical assistance and support with the computing cluster. J.B.O.Q. and C.A.V.T. acknowledge support from the Vicerrectoría de Investigaciones of the Universidad del Valle through Grant No. 71373. 

The trained models, predictions, and statistics extracted from the testing and extrapolation sets are hosted under the public GitHub repository: \\
\begin{center}
    \url{https://github.com/BayronO/MG-NECOLA}
\end{center}

\end{acknowledgments}

\bibliography{Biblio}{}
\bibliographystyle{aasjournalv7}

\end{document}